\documentclass[twocolumn,amsmath,amssymb,showpacs,prb,preprintnumbers]{revtex4}
\usepackage{epsfig}

\begin{document}
\newpage
\begin{abstract}
We present $^{13}$C NMR spin-lattice relaxation measurements (1/T$_1)$ in Na$%
_2$CsC$_{60}$ and Rb$_3$C$_{60}$ from 10 to 700~K. The large
temperature range of this measurement allow to define
unambiguously an increase of 1/T$_1$T with increasing temperature,
which is anomalous in a simple metallic picture, where the
Korringa law predicts 1/T$_1$T = cst. We attribute this increase
to the existence of an additional relaxation channel related to
singlet-triplet (ST) excitations of Jahn-Teller distorted
C$_{60}^{2-}$ and C$_{60}^{4-}$. These units are formed within the
metal on very short time scales (10$^{-14}$ sec) that do not imply
static charge segregation. We show that the amplitude of the ST
component depends directly on the density of states, which
indicates an interplay between metallic and molecular excitations.
Such an interaction is also revealed by the high temperature behavior of Na$%
_2$CsC$_{60}$ and CsC$_{60}$, that we then discuss. A divergence
between the behaviors of 1/T$_1$, the NMR shift and the ESR
susceptibility is interpreted as the result of a rapid increase of
the lifetime of the charge carriers, signaling a tendency to
charge localization. In our analysis, the particular stability of
C$_{60}^{2n-}$ is then a common feature of all known metallic
fullerides and allow to reconcile apparently contradicting
properties of these systems.
\end{abstract}
\newpage

\title{Persistence of molecular excitations in metallic fullerides and
their role in a possible metal to insulator transition at high
temperatures}

\author{V. Brouet, H. Alloul }

\affiliation{Laboratoire de Physique des Solides, Universite Paris-Sud, Bat 510
91405 Orsay (France)
}

\author{S. Garaj, L. Forr\'o}
\affiliation{
Laboratoire des solides semicristallins, IGA-Departement de
Physique, Ecole Polytechnique Federale de Lausanne, 1015 Lausanne
(switzerland)}

\date{\today} \newpage
\maketitle
\newpage

\newpage
\section{Introduction}

The properties of strongly correlated materials have been in the focus of
solid-state research for many years. To describe the competition between
Coulomb and kinetic energies, the one band Hubbard model is widely used.
However, many real systems exhibit orbital degeneracy, which adds a degree
of freedom that is not always taken into account. Fullerides are one such
example as the lowest unoccupied molecular orbital that forms the narrow t$%
_{1u}$ conduction band in the solid is triply degenerate. Together with the
strong electron-phonon coupling characteristic of these materials, this
leads to the possibility of Jahn-Teller distortions (JTD). Whereas
predictions for the JTD can be done quite accurately for a single molecule
\cite{Manini}, their role in the solid is less clear. The broadening of the
molecular levels should at first sight reduce the gain of energy associated
to JTD. As it is estimated to be of the same order of magnitude than the gap
opened by the JTD, the survival of these distortions in the solid can be
questioned.

Yet, JTD appear to play a crucial role in many different
fullerides. This paper concludes a serie of three papers devoted
to NMR studies of different stoichiometries of alkali fullerides,
where we have already seen evidence for such effects. It is
widely believed that JTD contribute to turn Na$_2$C$_{60}$ and A$_4$C$_{60}$%
, which should be metals in a band picture, into non-magnetic
insulators. In our first paper (called hereafter paper I
\cite{BrouetPartI}), we have presented NMR data supporting this
scenario and we refer the reader to references therein to the
various papers invoking JTD in the properties of A$_4$C$_{60}$.
Perhaps more surprisingly, they offer the most likely way for
explaining the unexpected properties of the cubic quenched (CQ)
phase of CsC$_{60}$, the only alkali cubic metallic fulleride
phase known so far besides A$_3$C$_{60}.$ In this phase, we have
shown in ref. \cite{BrouetPRL99} and paper II  \cite{BrouetPartII}
that the electronic properties are inhomogeneous on the local
scale because spin-singlets are trapped on about 10\% of the
C$_{60}$ balls. We believe that these singlets are stabilized by a
JTD of the C$_{60}$ ball which is energetically more favorable for
C$_{60}^{2-}$ than C$_{60}^{-}$.

\begin{figure}[tbp]
\centerline{ \epsfxsize=0.45 \textwidth{\epsfbox{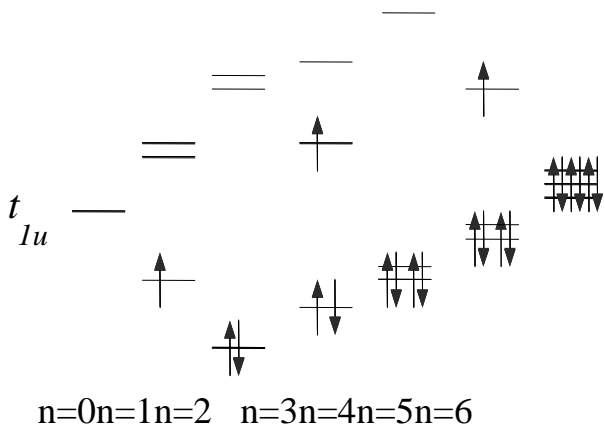}} }
\caption{Schematic representation of the splitting of the t$_{1u}$
levels as a function of the C$_{60}$ charge induced by Jahn-Teler
distortions (adapted from ref. \cite{Manini,Victoroff}). Only the
most stable JTD is represented.} \label{JT}
\end{figure}

More generally, a simplified view of the effect of JTD on the
t$_{1u}$ band as a function of its filling is sketched on
Fig.~\ref{JT}. The gain of energy obtained from the JTD is found
to be always larger for evenly charged C$_{60}.$ Then, JTD could
induce attractive interactions for odd
stoichiometries in order to promote the formation of the more stable C$%
_{60}^{2n-}$ (hence, the presence of spin-singlet in CQ
CsC$_{60}$), whereas for even stoichiometries, they would induce
repulsive interactions and favor localization (hence, the
insulating Na$_2$C$_{60}$ and A$_4$C$_{60}$). This suggests an
elegant way for rationalizing the different properties of even and
odd stoichiometries \cite{Manini,Heritier,Gunnarsson} and the
purpose of this paper is to determine to what extent this
framework could be relevant
for A$_3$C$_{60}$, extending ideas already presented in for the case of Na$%
_2 $CsC$_{60}$ \cite{BrouetPRL2001}.

Despite the possible survival of these unusual molecular properties in the
solid, there is a general consensus that the electronic properties of A$_3$C$%
_{60}$ can be understood in a rather conventional way, both for
the metallic and superconducting states \cite{GunnarssonSupra}. In
the first part of this paper, we show that there are however in
A$_3$C$_{60}$ deviations with respect to the conventional metallic
NMR behavior, that we attribute to an enhanced stability of
C$_{60}^{2-}$ and C$_{60}^{4-}$.\ The consequences of the
formation of such ``pairs'' on the properties of the metal is an
interesting issue to clarify. For this study, we choose two
examples of superconducting fullerides Rb$_3$C$_{60}$ (T$_c=29$~K)
and Na$_2$CsC$_{60}$ (T$_c=12$~K). While the structure of Rb$_3$C$
_{60}$ is face centered cubic ($fcc$, space group Fm$\overline{3}$m)
 through the whole temperature range \cite{StephensNature91}, Na$_2$CsC$%
_{60} $ undergoes an orientational transition from $fcc$ to simple
cubic ($sc$, space group Pa$\overline{3})$ around 300~K
\cite{PrassidesScience94}. In the second part of this paper,
we study the high temperature properties of two metallic fullerides where C$%
_{60}^{2-}$ have been detected, Na$_2$CsC$_{60}$ and CsC$_{60}$.
The behavior departs from the one of a metal and can be best
described by a progressive localization of the charge carriers
with increasing temperature. This is likely related to a change of
the role of the JTD as a function of temperature. However, the
most important conclusion of this study might be that it reveals
an overall similar behavior in these two compounds, which
simplifies greatly the understanding of fullerides by unifying
apparently conflicting properties. Up to now, CsC$_{60}$ was
thought to be metallic in its low temperature phase
\cite{KosakaPRB95} and insulating in its high temperature phase
\cite{TyckoPRB93}. The connection between these two behaviors was
not understood, which has forbidden so far a comparison with
A$_3$C$_{60}$ systems.

\section{Detection of molecular excitations through spin-lattice relaxation
measurements in superconducting fullerides}

In this part, we demonstrate that the deviation of 1/T$_1$T from
the simple metallic behavior can be convincingly attributed to
singlet-triplet distortions of JTD C$_{60}^{2-}$ and
C$_{60}^{4-}$. To establish this, we discard in the first part
other possible origin for this deviation. In the second part, we
give a quantitative analysis of the coexistence between metallic
relaxation and singlet-triplet excitations, which sets this model
on a firm basis.

\subsection{Origin of the non-Korringa relaxation}

\begin{figure}[tbp]
\centerline{ \epsfxsize=0.45 \textwidth{\epsfbox{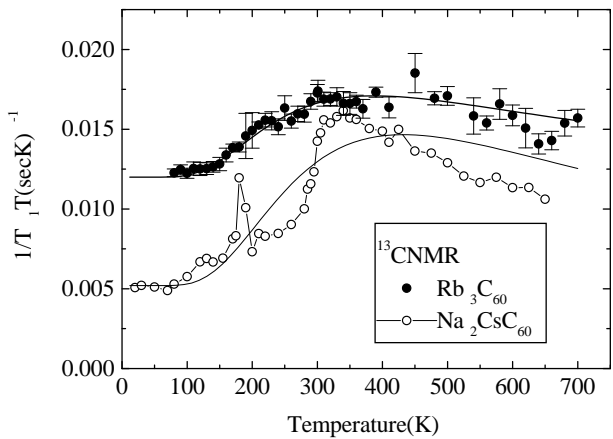}} }
\caption{ $^{13}$C NMR 1/T$_1$T as a function of the temperature
for Rb$_3$C$_{60}$ and Na$_2$CsC$_{60}$. The thick line
(Rb$_3$C$_{60}$) and thin line (Na$_2$CsC$_{60}$) are fits to the
Eq. \ref{twochannel} of the text. The parameters are displayed in
Table 1.} \label{A3C60}
\end{figure}

In a metal, the relaxation of nuclear spins is usually dominated
by their coupling with conduction electrons. This leads to a
simple dependence of the relaxation rate on the density of states
n(E$_f)$, known as the Korringa law \cite{Slichter}.
\begin{equation}
\label{korringa}\frac 1{T_1T}=\frac{\pi k_B}\hbar A^2n(E_f)^2
\end{equation}
where $A$ is the hyperfine coupling (in erg) between $^{13}$C and
conduction electrons. One anomaly in the NMR
behavior of A$_3$C$_{60}$ is that 1/T$_1$T frequently deviates from the 1/T$%
_1$T~=~cst law that one consequently expects. An increase of $^{13}$C NMR 1/T%
$_1$T with increasing temperatures has been reported for nearly all A$_3$C$%
_{60}$ compounds
\cite{TyckoPRL92,ManiwaJapan93,ManiwaPRB95,StengerPRL95},
sometimes with moderate values (30 \% in Rb$_3$C$_{60}$
\cite{TyckoPRL92}) or very large ones (200\% in Na$_2$CsC$_{60}$
\cite{ManiwaPRB95}). To probe these aspects further, we have taken
data in these two extreme cases in identical experimental
conditions, with a 7~T applied magnetic field and standard
saturation recovery pulse sequences, and over an extended
temperature range. The results are presented on Fig.~\ref{A3C60}.
In both compounds, the relaxation curves for the NMR magnetization
are not
exponential below 150~K, which causes {\it a priori} a problem for defining T%
$_1$ at low temperatures. However, we have shown in paper I that
the non-exponentiality is small and does not change between 10 and
150~K so that it cannot affect significantly these results.
Moreover, the region below 150~K is precisely the one where the
Korringa law is better obeyed. For the data presented here, T$_1$
has been defined below 150~K as the mean value of a double
exponential fit.

The narrow peak observed at 180~K in Na$_2$CsC$_{60}$ is due to a
coupling with slowing down molecular motions, as was also observed
in the isostructural Na$_2$C$_{60}$ \cite{BrouetPartI}. As this
shows that not only conduction electrons can contribute to the
relaxation in our case, one could wonder whether similar peaks
could be present (although not well resolved) in other temperature
ranges or in Rb$_3$C$_{60}$. If so, could they explain part of the
increase of 1/T$_1$T ? The large temperature range for the data
presented here allow to discard such a possibility, because these
peaks would be symmetric while the increase of 1/T$_1$T is
essentially step-like. We refer the reader to our paper I for a
detailed discussion of the characteristics of these molecular
motion peaks.

From Eq. \ref{korringa}, some temperature dependence of 1/T$_1$T
could be expected if n(E$_f$) is not strictly constant as expected
for a standard Pauli susceptibility. This could indeed happen, if
there are narrow features in the density of states near the Fermi
level. Those could be progressively disclosed as thermal expansion
increases the lattice constant with increasing temperature.
Because they are molecular solids, bound by weak Van der Waals
interactions, fullerides are actually very compressible materials
and the temperature dependence of n(E$_f$) must be seriously
considered. Some studies have concluded that it is sufficient to
explain the observed increase of 1/T$_1$T for certain compounds
\cite{StengerPRL95}. An obvious test to determine if the increase
of 1/T$_1$T is due to such an effect is to plot 1/T$_1$T together
with $\chi ^2$, as we have done for Na$_2$CsC$_{60}$ in ref. \cite
{BrouetPRL2001}. We have found that the temperature dependence of
$\chi ^2$ is much too small to explain that of 1/T$_1$T, so that
the deviation must be due, at
least in this case, to the presence of {\it an additional relaxation channel}%
.

We want to reinforce this conclusion here by taking advantage of
the comparison between Na$_2$CsC$_{60}\,$and Rb$_3$C$_{60}$, over
the large temperature range of the present experiments. The
accuracy of the data presented on Fig.~\ref{A3C60} at low
temperature establishes that the increase is not regular. While
1/T$_1$T is quite remarkably constant below 100~K
(Na$_2$CsC$_{60})$ or 150~K (Rb$_3$C$_{60}$), it then increases up
to
room temperature, where it decreases (Na$_2$CsC$_{60}$) or saturates (Rb$_3$C%
$_{60}$). The lattice contraction on the other hand is expected to
follow a smooth temperature dependence defined by the
compressibility of the materials. Furthermore, the compressibility
of $sc$ and $fcc$ phases are known to be quite similar
\cite{Samara}, so that the slightly different structure
 in the two compounds cannot explain such a difference. The thermal compressibility measured in Rb$%
_3 $C$_{60}$ ($\kappa =d(\ln a)/dT=3.10^{-5}$~K$^{-1}$
\cite{Zhou}) corresponds to an increase of the lattice parameter
$a$ by 0.16 {\AA} between 10 and 400~K, very close from the one
measured in Na$_2$CsC$_{60}$ (0.14 {\AA}
\cite{PrassidesScience94}). One could argue that even though the
temperature dependence of $a$ is the same, n(E$_f$) might display
quite different variations with $a$ in $sc$ or $fcc$ materials
\cite{YldirimSScom95}. This idea has been proposed to explain the
different variation of $T_c$ with $a$ in Na$_2$AC$_{60}$ ($sc$) or
A$_3$C$_{60}$ ($fcc$) materials (with A = K, Rb, Cs) because it is
generally assumed that T$_c$ depends on n(E$_f$) in a
straightforward way according to BCS theory. However, the
variation of T$_c$
with pressure was later found the same in $fcc$ and $sc$ phases \cite{BrownPRB99}%
, so that the different variation of T$_c$ is not related to
n(E$_f$) but to an alkali effect. The comparison between
Na$_2$CsC$_{60}$ and Rb$_3$C$_{60}$ is then legitimate and
presents the advantage to cover a large variation in density of
states. An even more serious drawback with such an explanation is
that our study evidences a much larger increase in
Na$_2$CsC$_{60}$ than in Rb$_3$C$_{60},$ although a direct
measurement of the susceptibility below 300~K shows the opposite
behavior \cite{RobertPRB98}.

\subsection{Singlet-triplet excitations in A$_3$C$_{60}$}

In Na$_2$CsC$_{60}$, the origin of the increase of 1/T$_1$T was
suggested by the comparison with Na$_2$C$_{60},$ which exhibits a
very similar relaxation behavior at high T despite it is
insulating (see \cite {BrouetPRL2001} or Fig.~\ref{T1ccl}). In
this latter compound, the
relaxation is due to singlet-triplet transitions between different JTD of a C%
$_{60}^{2-}.$ Theoretical calculations have shown that for an isolated C$%
_{60}^{2-}$, a JTD with a triplet ground state lies above the
singlet JTD represented in Fig.~\ref{JT} by E$_a$=140~meV, so that
triplet states can be
thermally populated \cite{Manini}. Therefore, we have suggested that {\it %
singlet-triplet excitations persist in Na$_2$CsC$_{60}$} because C$%
_{60}^{2-} $ and C$_{60}^{4-}$ are formed there on very short time
scales (about 10$^{-14}$~sec), which creates an additional
relaxation channel explaining the increase of 1/T$_1$T at high
temperature. We now investigate if such an explanation could
describe the evolution of the relaxation behavior between
Na$_2$CsC$_{60}$ and Rb$_3$C$_{60}$.

The data led us to suggest that the relaxation can be divided
between two relaxation channels, a metallic Korringa-like channel
and a molecular channel corresponding to localized singlet-triplet
excitations. For this latter term, we assume that the imaginary
part of the susceptibility, the quantity measured by 1/T$_1$ at
the nuclear Larmor frequency (80 MHz for $^{13}$C in this
measurement), can be described by a Lorentzian with a width
1/$\tau$, where $\tau$ is a lifetime characterizing the electronic
excitations. This is is a usual assumption for localized magnetic
moments, which yields for small $\omega$, 1/T$_1\propto \chi \tau
$, where $\chi$ is the static paramagnetic susceptibility of the
localized moment \cite{White}. This leads to the following
expression.
\begin{equation}
\label{twochannel}\frac 1{T_1T}=\frac{k_B}\hbar A^2\text{ }\left(
\pi n(E_f)^2+\frac{\chi _{_{ST}}}{\mu _B^2}\frac{\tau _{st}}\hbar
\right)
\end{equation}
where $\tau _{st}$ is the characteristic lifetime of a triplet state and $%
\chi _{_{ST}}$ an activated susceptibility describing the
population of triplet states. The separation between the two terms
is quite arbitrary, as the same electrons participate to both
terms. Such a phenomenological
decomposition should nevertheless capture the essential points of the physics as long as $%
\tau _{st}$ is shorter than the lifetime $\tau _p$ of a
C$_{60}^{2n-}$ in the metal. Some correlation between the
parameters describing the ``localized'' states (like $\tau _{st}
$) and the extended carriers states (like n(E$_f$)) can be
expected as a result of this situation.

\begin{figure}[tbp]
\centerline{ \epsfxsize=0.45 \textwidth{\epsfbox{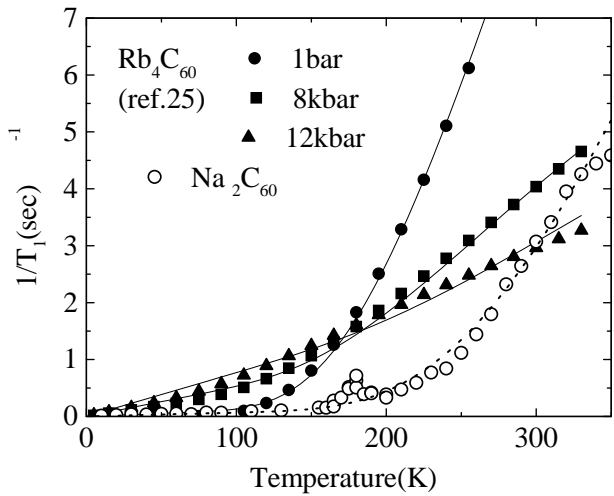}} }
\caption{$^{13}$C NMR 1/T$_1$ in Rb$_4$C$_{60}$ for different
applied pressure (from ref. \cite{Kerkoud}) and Na$_2$C$_{60}$.
The curves are fitted to Eq.~\ref{twochannel} with parameters
displayed in Table~1. They assume that the relaxation is the sum
of a linear metallic component and an activated molecular one with
fixed ratio.} \label{simultriplet}
\end{figure}

A source of inspiration for this correlation is the evolution of Rb$_4$C$%
_{60}$ from an insulator to a metal with applied pressure
\cite{Kerkoud}. This transition is not sharp, but a linear
metallic term appears with increasing pressure {\it that coexists}
with the localized activated one, as replotted on
Fig.~\ref{simultriplet}. As the metallic term grows, the molecular
one can still be described by the same model, but the value of the
gap and $\tau _{st}$ have to be reduced. This legitimates the
assumption of coexisting molecular and metallic excitations, which
is also observed in Na$_2$C$_{60}$ \cite{BrouetPartI}. We believe
that this property is actually essential to understand the
originality of fullerides, in which molecular features are
retained because the large electronic correlations forces one
electron to spend a ``long'' time on each molecule before being
transferred to the next.

The behavior in Rb$_4$C$_{60}$ tells us that $\tau _{st}$ directly
depends on n(E$_f$). This suggests that {\it the triplet states
are relaxed by conduction electrons}. Assuming that the triplet
state can be treated as a local impurity on the timescale of its
existence, we can use relaxation laws observed for localized
impurities in a metal, that are analogous to a Korringa-like
process \cite{White}.
\begin{equation}
\label{tauST}\frac 1{\tau _{st}}=\frac \pi \hbar \text{
}k_BT\text{ \ }J^2 \text{ }n(E_f)^2
\end{equation}
where J would be the coupling between the triplet and conduction
electrons.

\begin{table*}

\begin{ruledtabular}

\begin{tabular}{ccccccc}

 &\multicolumn{3}{c}{Rb$_4$C$_{60}$}& Na$_2$C$_{60}$&   Na$_2$CsC$_{60}$&    Rb$_3$C$_{60}$\\
   &1 bar&8kbar&12kbar&\\ \hline

$\chi _m$ (10$^{-4}$emu/mol)   &      1   &   3.1 &    3.7  &   1.3 &   3.1&   4.7\\
 $\Delta$ (meV)     &    90  &      80 &100&   125  &   85&   75\\
$\tau _m=\hbar n(E_f)$ (10$^{-15}$sec)&         &   6 &    7.2  &   &   6 &   9\\
 $\tau_{st}$(300K) (10$^{-15}$ sec)&     35.3       &    3.7  &  2.6  &   21   &   3.7&   1.6\\
\end{tabular}
\end{ruledtabular}

\caption{\label{tab:table3}Parameters used in the fits shown on
Fig.~2 and Fig.~3 based on Equation 2, 3 and 4. $\chi _m$ is
deduced from n(E$_f$) and Eq. 1 assuming an hyperfine coupling
A=4.10$^{-20}$ erg. The absolute values of $\tau_{st}$ should only
be taken as an order of magnitude because it is somewhat model
dependent but the qualitative variation between different
compounds is meaningful, as well as the condition $\tau_{st}$$\leq
$ $\tau_{m}$ which legitimates the use of this model for the
metals.}

\end{table*}

To test this model, which reduces the number of free parameters in
Eq. \ref {twochannel}, we have tried to reproduce the pressure
evolution of the two terms in Rb$_4$C$_{60}$. To modelize the
susceptibility associated with singlet-triplet excitations, we
choose
\begin{equation}
\label{chiST}\chi _{st}=\frac{8\mu _B^2}{k_BT}*\frac{exp(-\Delta /T)}{%
2+3\exp (-\Delta /T)}
\end{equation}
which was found in reasonable agreement with the susceptibility of Na$_2$C$%
_{60}$ \cite{BrouetPartI}. The coupling $J$ is treated as a shared
parameter between the three sets of data and the best fits are
found for $J=10^{-13}$ erg. Fits using equations \ref{twochannel},
\ref{tauST} and \ref{chiST} with this value for J are shown on
Fig.~\ref{simultriplet}. They are in fair agreement with the data,
which supports this model. The density of states and the value of
the gap used in each case are displayed in Table 1. This model can
be applied as well to Na$_2$C$_{60}.$ Fig.~\ref{simultriplet}
shows that, with a larger gap value already noticed previously
\cite {BrouetPRL2001} but keeping the same value for J, the data
are also well reproduced.

Does this model also apply to the case of A$_3$C$_{60}$ ? Eq. 3
implies that the increase of 1/T$_1$T due to the localized term
should show up more clearly when n(E$_f$) is small, which is in
qualitative agreement with the evolution of 1/T$_1$T between
Na$_2$CsC$_{60}$ and Rb$_3$C$_{60}$. Remarkably, the previous
fitting procedure {\it with the same J value} gives a correct
order of magnitude for the increase of 1/T$_1$T in both compounds.
In this model, the magnitude of the high temperature increase of
1/T$_1$T is uniquely fixed by its value at low temperature. The
best fits displayed on Fig.~\ref{A3C60} with the parameters given
in table 1 show an excellent agreement in the case of
Rb$_3$C$_{60}$ but is poorer for Na$_2$CsC$_{60}.$ In this latter
compound, it seems that a different regime is relevant for high
temperatures, which will be discussed in the next section. Even
below 300 K, the increase of 1/T$_1$T cannot be ascribed to a
single gap value. Similar problems were encountered in the case of
Na$_2$C$_{60}$, which led us to suggest in paper I that the
equilibrium between different JTD could be modified by changes in
details of the structure as a function of temperature.

To check the consistency of our analysis, the triplet relaxation
time $\tau _{st}$ should be compared to the lifetime $\tau _p$ of
the C$_{60}^{2n-}$. This latter time can only be longer that the
average time spent by one electron in the vicinity of a C$_{60}$
ball $\tau _m$. A lower bound for $\tau _m$ can be estimated by
the time required for an electron to travel at the Fermi velocity
from one ball to the other, $\tau _m$= a / v$_f\approx \hbar n(E_f$%
). This is calculated in Table 1 and the comparison with $\tau
_{st}$ deduced from the fits shows that they have similar order of
magnitude at room temperature, so that this analysis is
consistent. We note that we have used here the susceptibility
given by Eq. \ref{chiST}, although one could assume that only a
fraction of it, corresponding to the actual concentration of
C$_{60}^{2n-}$ in A$_3$C$_{60}$, should be used. This would give
slightly longer value for $\tau _{st}$.

Therefore, this simple model gives an overall satisfying
description of the relaxation behavior in these compounds. The
correlation between the molecular and metallic term might however
be even more intricate. We have assumed that the JTD C$_{60}^{2-}$
can be considered as an isolated molecular entity, even if it is
restricted to a limited timescale. The fact that the gap value is
not fixed, but tends to decrease with increasing n(E$_f$) in
compounds with identical structures like Na$_2$C$_{60}$ and
Na$_2$CsC$_{60},$ shows explicitely that there is a further
correlation between the two terms. $\Delta $ is not purely a
molecular value and a more sophisticated description, introducing
for example a screening of the C$_{60}^{2n-}$ by conduction
electrons, would be required to describe completely this behavior.

In addition, we have mainly tried to explain so far the increase
of 1/T$_1$T which roughly takes place between 200 and 300 K. At
higher temperatures, Eq. \ref{tauST} predicts a decrease of the
lifetime of the triplet states as 1/T. This could change the
balance between the two relaxation channels as a function of
temperature. The simple estimate that we have used for $\tau _m$
is also likely to break down at high temperature because it
requires that the mean free path for the electronic motion $l$ is
longer than $a$. On the contrary, it is known that very small
values for the mean free path are deduced from resistivity
measurements for fullerides at high temperature, like 1-2 {\AA} at
500~K in Rb$_3$C$_{60},$ \cite{HebardPRB93}. Therefore, we could
expect complications at high temperature, that might reveal, on
the other hand, valuable information on the charge transport in
these materials and we turn to this topic in the next section.

\section{Are ``metallic fullerides'' metallic up to high temperatures ?}

Some anomalies in the high temperature behavior of fullerides have
been noticed for a long time. In Rb$_3$C$_{60},$ the photoemission
spectra at
high temperature does not display a clear Fermi edge \cite{Golden}. In Na$_2$%
CsC$_{60}$, optical conductivity measurements indicate a
disappearance of the Drude-like peak above 300 K \cite{Cegar}. All
this motivates us to investigate the high temperature region in
more details and we indeed report here two cases, Na$_2$CsC$_{60}$
and CsC$_{60}$, where the high temperature behaviors are far from
that of a simple metal and are even suggestive of a localization
of the charge carriers. For all the data reported here, we have
checked that the behavior is fully reversible with temperature,
ensuring that there is no deterioration of the sample quality
and/or stoichiometrie.

\subsection{Na$_2$CsC$_{60}$}

\subsubsection{progressive localization of the charge carriers inferred from
1/T$_1$ behavior}

Fig.~\ref{A3C60} shows that 1/T$_1$T decreases regularly from 300
to 700 K in the $fcc$ phase of Na$_2$CsC$_{60}$. This is anomalous
as among the two terms identified in the relaxation so far, the
metallic and the singlet-triplet components, the first one is
expected to yield a constant value and the second one an increase
of 1/T$_1$T or a slight decrease
depending on the precise value of the gap (see the fit of Fig.~2). In Rb$_3$C%
$_{60},$ there might be a similar anomaly above 500 K, where
1/T$_1$T is somewhat lower than the fitted curve, but this cannot
be concluded unambiguously within experimental accuracy and in the
following we focus on Na$_2$CsC$_{60}.$

As the use of 1/T$_1$T might introduce an artificial bias by
emphasizing the metallic component, we have replotted on
Fig.~\ref{InvT1TCsCar}, 1/T$_1$ for $^{13}$C NMR, together with
the result from $^{133}$Cs NMR, which displays essentially the
same temperature dependence, except for the molecular motion peak.
In our case, the interpretation of these data is not
straightforward as we do not know how to separate the metallic
term from the ST component. Within the metallic framework of the
Korringa law, a decrease of 1/T$_1$T would be assigned to a
reduction in the density of states signaling the onset of a broad
metal to insulator transition. If it is a change in the
singlet-triplet component, it could be due to modifications of the
apparent value of the gap as was observed in Na$_2$C$_{60}.$ To
find a way out of this problem, a comparison with the ESR
susceptibility shown in the inset of Fig.~\ref{InvT1TCsCar} is
helpful. In the susceptibility, both the metallic and
singlet-triplet terms are also present but with different
ratios as they are not weighted by the lifetime of the excitations, like 1/T$%
_1$ is. $\chi $ is dominated at low T by a constant Pauli
component and then increases up to 350 K, in our opinion mainly as
a result of singlet-triplet excitations. Like 1/T$_1$T, it is
clearly anomalous above room temperature, where it starts
decreasing sharply. This is very different from the behavior of
$\chi $ in Na$_2$C$_{60}$, where it keeps increasing at high
temperature, which tells us that a modification of the ST
component alone cannot account for this result. Above 350~K, we
observe a strong divergence in behavior between $\chi $ and
1/T$_1,$ $\chi $ falls far below the metallic value, whereas
1/T$_1$T is still far above. This can only be understood if {\it
the lifetime of the excitations diverges above room temperature}.
This reminds of a situation observed at the transition between
solid and liquid in X$_2$Te$_3$ (X=In, Ga, Sb) \cite{Warren}. Upon
melting, the increasing disorder induces a progressive
localization of the charge carriers that leads to a decrease of
the Pauli susceptibility, due to the reduction of the carrier
density. At the same time, 1/T$_1$T is found to increase, as the
lifetimes of the excitations increases rapidly. To account for
this effect n(E$_f$)$^2$ in Eq. 2 has been replaced by the authors
of ref \cite{Warren}, by n(E$_f$)$\tau _m/\hbar ,$ where $\tau _m$
is now free to deviate from the metallic value $\hbar n(E_f$). In
our case, both $\tau _m$ and/or $\tau _{st} $ could be responsible
for the increase of 1/T$_1$T. However, in both cases, we can
conclude that there is {\it a progressive localization of charge
carriers. }

\subsubsection{nature of the localized charge carriers inferred from the
static NMR alkali spectra}

\begin{figure}[tbp]
\centerline{ \epsfxsize=0.45 \textwidth{\epsfbox{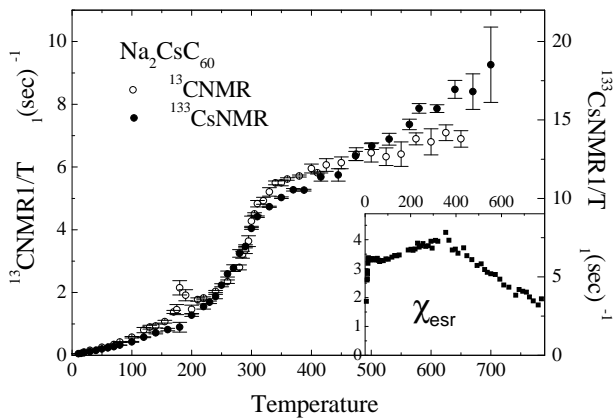}} }
\caption{ 1/T$_1$ for $^{13}$C and $^{133}$Cs in Na$_2$CsC$_{60}$
as a function of temperature. For $^{133}$Cs, the recovery curves
are exponential through the whole temperature range. Inset : ESR
susceptibility as a function of temperature measured on a sample
from the same batch.} \label{InvT1TCsCar}
\end{figure}

If the localization process went as far as to correspond to a
static charge separation, we could expect to observe changes in
the NMR spectra as a
result of inequivalent alkali sites neighboring a different number of C$%
_{60}^{2n-}$. This is the case in CQ CsC$_{60}$ where a splitting
of the Cs
spectrum is observed at low temperature, corresponding to the localization of 10\% C$%
_{60}^{2-}\,$, as discussed in details in paper II. In
Na$_2$CsC$_{60,}$ static charge separation can be ruled out
because $^{133}$Cs and $^{23}$Na spectra consists of one
featureless narrow line (with the exception of the Na ``T'~''
line). This is consistent with the value given previously for
$\tau _{st}$ which is much shorter than the NMR time scale (a few
ms corresponding to the inverse width of the spectrum), so that a
motional narrowing of the spectra should still take place even if
$\tau _{st}$ increases by many order of magnitudes.

\begin{figure}[tbp]
\centerline{ \epsfxsize=0.45 \textwidth{\epsfbox{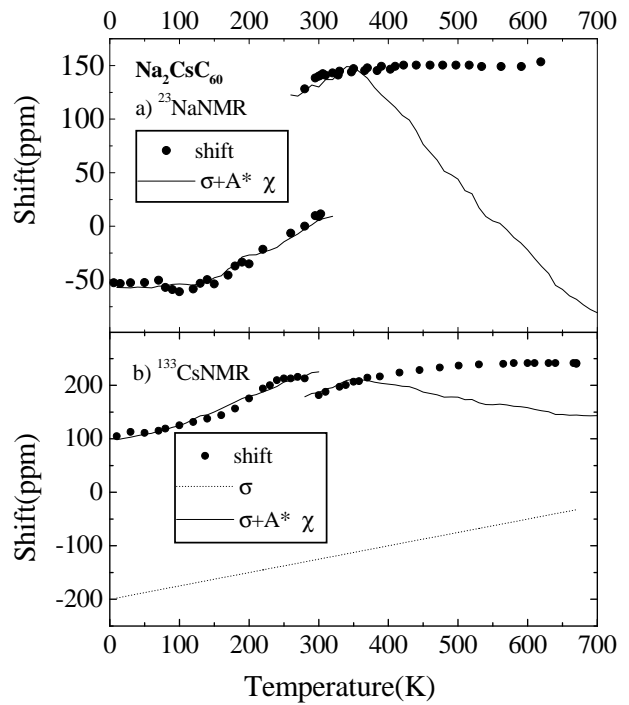}} }
\caption{Shift for $^{23}$Na (part a) and $^{133}$Cs (part b) in
Na$_2$CsC$_{60}$. The discontinuity at 300~K is due to the
structural transition from the orientationally ordered $sc$ phase
(T$<300$~K) to the $fcc$ one. $^{23}$Na (resp. $^{133}$Cs) shifts
are measured with respect to a diluted NaCl (resp. CsCl) solution.
The lines are comparison with the behavior of the susceptibility
assuming values of hyperfine shift and chemical reference
discussed in the text.} \label{shiftNa2CsHT}
\end{figure}

There are however clear anomalies in the alkali NMR which are
revealed when comparing the shifts $K$ for $^{133}$Cs and
$^{23}$Na lines, presented on Fig.~\ref{shiftNa2CsHT}, together
with the ESR susceptibility. A simple scaling between $K$ and
$\chi $ is expected, $K=\sigma +A\chi $, where A is the hyperfine
coupling and $\sigma $ a reference chemical shift. On Fig.~\ref
{shiftNa2CsHT}, it can be seen that both nuclei sense an increase
of the susceptibility between 150 K and 300 K, but they do not
reproduce similarly the decrease seen by ESR above 300 K. This
also suggests, although in a puzzling way, that the properties of
the material are changing above room temperature. Let us now
discuss quantitatively the behavior of the shift for the two
nuclei.

\paragraph{$^{23}$Na NMR}

In the $sc$ phase, we can scale the $^{23}$Na shift and the
susceptibility, as shown by the line of Fig.~\ref{shiftNa2CsHT}a
by using $\sigma =-300$ ppm and A = 4500~Oe/$\mu _B$. It is
interesting to compare these numbers with
those found for $^{23}$Na NMR in paper I for the isostructural Na$_2$C$%
_{60},$ $\sigma $=-65 ppm and $A$= 2300~Oe/$\mu _B$. The difference for $%
\sigma $ is quite large and not expected as $\sigma $ essentially
depends on the diamagnetism of the Na$^{+}$ filled orbitals which
should not change much. To avoid this problem, it might be more
realistic to assume that the hyperfine coupling to the metallic
component is {\it smaller }than the one
to localized excitations, maybe because Na moves a little bit towards the C$%
_{60}^{2-}$ or C$_{60}^{4-},$ which increases the hyperfine
coupling for
this term. This automatically gives a value of $\sigma $ closer to that of Na%
$_2$C$_{60}$, as the shift is dominated at low T by the metallic
term.

Through the $sc$-$fcc$ transition at 310 K, we have observed in Na$_2$C$%
_{60}$ an increase by 30\% of the hyperfine coupling probably
associated to the change of structural environment of Na (see
paper I). This is also clearly present in Na$_2$CsC$_{60}$. As a
guide to the eye, we have plotted the variation of $K$ that would
be expected for $^{23}$Na NMR from the one of $\chi $ assuming no
change for $\sigma $ and a continuous value for $\chi .$ This only
emphasizes what was immediately clear, $K$ does not follow the
static susceptibility at high temperature, it is nearly constant
while $\chi $ drops by a factor 2. Relying on the previous
findings i)~the characteristic lifetime of the localized
excitations starts to increase above 300 K and ii)~the hyperfine
coupling to the localized excitations is particularly strong, we
would like to conclude that the reason for the different behavior
of K and $\chi $ at high temperature is due to {\it an increase of
the hyperfine coupling associated with the progressive
localization of the carriers}. Indeed, we know from CQ-CsC$_{60},$
studied in paper II, that the coupling can be very strong with
localized singlets in a metallic environment. This enhancement
disappears if all the balls are equivalent and bear a singlet,
like in Na$_2$C$_{60}$, because Na has no reason to move towards a
particular ball.

\paragraph{$^{133}$Cs NMR}

The $^{133}$Cs shift presents a similar behavior but it displays a
significant temperature variation below 100 K, which does not
appear in the ESR data. Such a linear temperature dependence of
the shift has already been observed in the NMR of various alkali
in fullerides \cite {AlloulPhysica,Stenger,ZimmerEurophysics}. Its
origin is not well understood, but as it is not related to the
electronic susceptibility \cite {AlloulPhysica}, it is probably of
orbital origin and we will include this linear variation in the
chemical shift $\sigma $. Choosing $\sigma =-200+0.25$~$T,$ as
indicated on Fig.~\ref{shiftNa2CsHT} by the dotted line, allows to
extract an additional increase, clearly present
between 150K and 300K which scales with the susceptibility with A= 5000 Oe/$%
\mu _B$. As done for $^{23}$Na, we then extrapolate the variation
of $K$ at high temperatures assuming that $\sigma $ and $\chi $ do
not change, which requires here a reduction of A in the $fcc$
phase to 4500 Oe/$\mu _B$ \cite {rq1}. Hereagain, there is a
significative difference between K and $\chi ,$ although less
dramatic than for Na. We also observe that the scaling between K
and $\chi $ is not lost at the structural transition but at
somewhat higher temperatures T$\approx $~400 K, making it unlikely
that the structural transition could be responsible for the change
of the electronic properties of Na$_2$CsC$_{60}.$ If the
divergence between K and $\chi $ is to be attributed again to an
increase of the hyperfine coupling, this yields A=6000~Oe/$\mu _B$
at 650K, which corresponds to a 25 \% increase from 350~K to
650~K. Such an increase should also enhance 1/T$_1,$ proportional
to A$^2, $ by a factor 1.5. On Fig.~\ref{InvT1TCsCar}, we see that
$^{133}$Cs 1/T$_1$ indeed increases a little bit more rapidly than
$^{13}$C 1/T$_1$ at high temperatures. As $^{13}$C cannot move,
its hyperfine coupling can be used as
a reference for the ``unenhanced'' value of 1/T$_1.$ This predicts 1/T$_1$%
=13~sec$^{-1}$ instead of the observed 1/T$_1$=17~sec$^{-1},$
which is close from a factor 1.5 and supports the idea of an
enhancement of the hyperfine coupling \cite{rq2}.\medskip

In summary, many anomalies appear above 400~K in Na$_2$CsC$_{60}$,
which gives evidence for a change in the electronic properties.
Although, the data are conflicting at first sight, because
different probes measure apparently different behaviors, we
propose that this can be understood assuming that we enter an
intermediate regime between metallic and localized behavior. In
this regime, 1/T$_1$ is dominated by the increase of the lifetime
of the charge carriers around a particular C$_{60}$ ball, which
also modulates the value of the alkali hyperfine couplings. We
cannot decide unambiguously whether the charge carriers are
localized as C$_{60}^{3-}$ or C$_{60}^{2n-}$. Both are consistent
with the 1/T$_1$ measurements. The change in the alkali hyperfine
coupling somewhat favors C$_{60}^{2n-}$, as we could expect a
larger motion of the alkali towards the differently charged
C$_{60}$. The idea that a motion of the alkali ions, especially of
the light Na ion, is playing an important role in the high
temperature properties of this phase was already suggested in
\cite{Cegar}. However, we cannot rule out that the appearance of a
localized charge is sufficient to trigger such a motion and modify
the alkali hyperfine coupling regardless of the charge value.

Because we have detected in the previous part of this paper an
important role of JTD in the metallic phase, it seems reasonable
to associate the progressive localization at high temperatures
with a conflict between the stabilization of JTD and the
delocalization of the charge carriers. We can for example
speculate that the population of many different distortions at
high temperatures introduces disorder for the electronic system,
which hinders the electronic motion. These electrons could then be
localized as C$_{60}^{3-}$ or C$_{60}^{2n-}$ depending on the
precise balance between Coulomb repulsion and Jahn-Teller
attraction, but, more importantly, this behavior reveals by defect
that a cooperation is required between JTD and electronic motion
to obtain the metallic state. This agrees with the idea that JTD,
at last at low temperatures, help the systems with odd
stoichiometries to become metallic by stabilizing the
C$_{60}^{2n-}$, which have to be formed as a byproduct of
electronic jumps.

\subsection{The high temperature cubic phase of CsC$_{60}$}

\begin{figure}[tbp]
\centerline{ \epsfxsize=0.45 \textwidth{\epsfbox{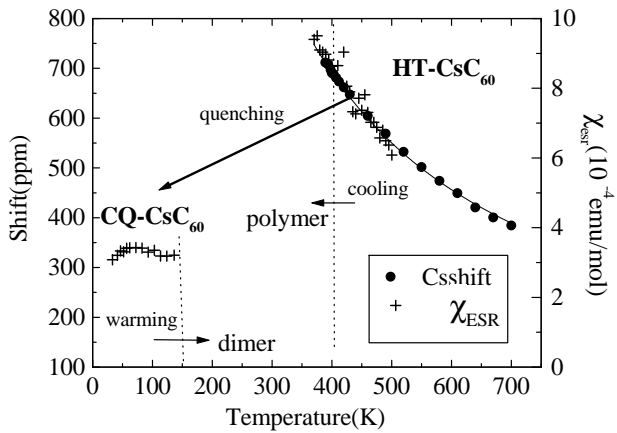}} }
\caption{$^{133}$Cs NMR shift in the high temperature cubic phase
of CsC$_{60}$ (black points), the line is a fit to a Curie law.
There are two cubic phases in CsC$_{60}$, a cubic quenched (CQ)
metastable phase is stable only for T$<130$~K, and a high
temperature one (T$>$380~K). In these two phases, crosses show the
behavior of the ESR  susceptibility to illustrate the change in
electronic properties.} \label{Csshift}
\end{figure}

\begin{figure}[tbp]
\centerline{ \epsfxsize=0.45 \textwidth{\epsfbox{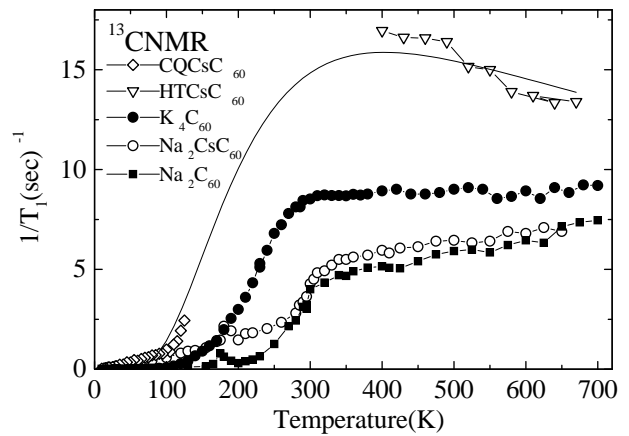}} }
\caption{Comparison of $^{13}$C NMR 1/T$_1$ in different
fullerides showing a similar increase up to 300~K that we
attribute to singlet-triplet excitations.  The line connecting the
cubic quenched (CQ) phase of CsC$_{60}$ and the high temperature
(HT) one is a fit to Eq.~\ref{twochannel} given in the text.}
\label{T1ccl}
\end{figure}

Another very anomalous high-temperature phase is the one of
CsC$_{60}$. It appears to have a Curie-like susceptibility
\cite{Chauvet}, the $^{133}$Cs NMR shift also follows a Curie law
within experimental accuracy and 1/T$_1$
is nearly constant as expected for a paramagnetic insulator \cite{TyckoPRB93}%
. All of this was taken as evidence for electronic localization.
As for NMR, we have checked that this behavior holds up to 700~K
for the $^{133}$Cs shift (see Fig.~\ref{Csshift}) and $^{13}$C
1/T$_1$ (see Fig.~\ref{T1ccl}, actually T$_1^{-1}$ is not really
constant but decreases by 25~\% from 400~K to 700~K).

Below 350~K, a structural transition takes place to a polymerized
phase with very different properties. However, the cubic phase can
be studied at low temperatures (below 130~K) by quenching the high
temperature phase {\it and it seems to be metallic}. A Pauli
susceptibility \cite{KosakaPRB95} and a Korringa law for $^{13}$C
NMR above 50~K \cite{BrouetPRL99} are observed. The behavior of
the susceptibility in the two phases are reported on
Fig.~\ref{Csshift} to illustrate this point. The difference is
very puzzling, as the structure is the same except for the
orientational order (the CQ phase of CsC$_{60}$ is $sc$
\cite{Lappas}, while the high temperature phase is $fcc$). If this
structural transition were to play a role in the conductivity, the
high temperature phase, which is structurally equivalent to metallic A$_3$C$%
_{60},$ should be the more directly comparable to other
fullerides, whereas it is this phase that displays ``anomalous''
properties. This distinct behavior in the two phases casts doubt
on whether 1 electron in the t$_{1u}$ band yields a metallic or
insulating behavior.

On the other hand, it appears now quite similar to that of
Na$_2$CsC$_{60}$ and it seems less difficult to connect the
properties of the two CsC$_{60}$ phases. Na$_2$CsC$_{60}$ also
resembles a good metal at low temperature and, for some
properties, particularly the ESR ``Curie-like'' susceptibility, an
insulator above 300~K. If the data between 100 and 300~K were missing for Na$%
_2$CsC$_{60}$, one would also conclude that we are dealing with
two entirely different phases. We have seen through this paper
that including singlet-triplet excitations of JTD balls offers a
convincing way to explain this apparent discrepancy, and the same
could be true for CsC$_{60}.$ The underlying similarity is perhaps
best illustrated by looking at the temperature dependence of
1/T$_1$ for $^{13}$C in various phases, as we do on
Fig.~\ref{T1ccl}. In this context, it is tempting to interpret the
gap between the values of 1/T$_1$ in the CQ phase and those of the
high temperature phase by an increase due to a singlet-triplet
component. In CQ CsC$_{60}$, just
before the transition to the dimer phase, we do observe an increase of 1/T$%
_1,$ which might correspond to such a component. Adjusting a fit
to Eq. \ref {twochannel} to these points and the high temperature
value gives the line sketched on Fig.~\ref{T1ccl}. This yields a
smaller gap $\Delta =50$~meV
than for the other compounds, but this is consistent with the fact that 1/T$%
_1$ decreases at high temperatures. Indeed, Fig.~\ref{T1ccl} shows
a systematic relation between the value of the gap and the
behavior at high temperature: the lower the gap, the higher
1/T$_1$ and the smaller (eventually negative) the slope of its
variation at high temperature. Qualitatively, this is not
surprising as, with a small gap, the susceptibility is dominated
at high temperature by the Curie law for the triplet states (see
Eq. \ref{chiST}).\medskip

Putting these data together then gives a fairly good understanding of the CsC%
$_{60}$ system. We know from our study of CQ CsC$_{60}$ that JTD
C$_{60}^{2-} $ are formed at low temperatures but, presumably
because there is no symmetric JTD for neutral C$_{60},$ they get
localized on a small fraction of the C$_{60}$ balls, contrary to
C$_{60}^{2n-}$ in superconductors. The
singlets start to move as the temperature increases and the lifetime of a C$%
_{60}^{2-}$ in the metal decreases exponentially as shown in paper
II. We can speculate that, if it were stable, the properties of
cubic CsC$_{60}$ would be very similar to those of a metal like
Na$_2$CsC$_{60}$ between 100~K and 300~K and would evolve
gradually towards an almost insulating state at high temperatures,
which is the one observed above 400~K.

\section{Conclusion}

To conclude, we have proposed in this paper that the metallic
character of fullerides with an odd number of electrons per
C$_{60}$ {\it is driven by the formation of pairs of electrons on
very short time scales stabilized by Jahn-Teller distortions}. We
argue that the deviation of the NMR 1/T$_1$T from the canonical
``metallic'' Korringa law observed in many A$_3$C$_{60}$ systems
can be convincingly attributed to the presence of singlet-triplet
excitations of these electronic pairs. By studying two
superconducting metals with very different density of states, we
evidence a dependence of the characteristic time $\tau _{st}$ of
the triplet state for these pairs on n(E$_f$) that makes the
detection of singlet-triplet excitations easier in compounds with
small density of states. This explains why the effect has not be
clearly identified up to now, the molecular excitations being
masked in the most heavily studied compounds like K$_3$C$_{60}$
and Rb$_3$C$_{60}$.

We have then presented new data at high temperatures that show a
clear evolution of the behavior of the electronic properties in at
least two compounds Na$_2$CsC$_{60}$ and CsC$_{60}$. In the first
case, we interpret the breakdown of the scaling between various
quantities at about 400~K as the onset of a progressive
localization of the charge carriers. We suggest that this
localization is due to a change of the status of JTD in the metal
as the temperature increases. The localized charge carriers could
either be the electronic pairs formed via JTD or independent
electrons, depending on the precise nature of the high temperature
evolution. A similar, probably slightly more efficient,
localization takes place in CsC$_{60}$. Although this phase was
already believed to be insulating, seeing this as a metal to
insulator transition allow to connect its properties to those of
the low temperature metallic phase. This adds greatly to the
understanding of these systems, as all metallic phases can now be
described within the same framework. JTD appear as a key
ingredient in this behavior, which makes it likely that the
difference between even and odd stoichiometries, respectively
insulating and metallic, is due to the sensitivity of Jahn-Teller
distortions to the parity of the C$_{60}$ charge. As recalled in
Fig. 1, JTD are always a little bit more stable for even parity.

This work also raise new questions on the nature of charge transport in A$_3$%
C$_{60}.$ If the enhanced lifetime of C$_{60}^{2-}$ and
C$_{60}^{4-}$ is a general feature of A$_3$C$_{60},$ it is of
course important to examine their role within the metal. The role
that we assigned here to Jahn-Teller distortion is an example of
how electronic properties are coupled to phonons and could be a
first step towards the concept of polarons. However, we do not
know how fast the JTD can adjust itself to a change of the charge
of the molecule, produced by jumps of electrons from one ball to
the other. The idea of a polaron would imply that one electron
moves with a given JTD. In our case, we could as well assume that
a JTD is fixed for a given molecule, regardless of its charge,
which only increases the lifetime of favorable JTD C$_{60}^{n-}$
configurations (namely n even). The lack of experimental knowledge
about the time scale for the Jahn-Teller distortion, relatively to
the motion of electrons, makes it difficult to discriminate
between these possibilities. Nevertheless, some correlation
between them is required to establish a coherent band-like charge
transport. This could be
destroyed with increasing temperature resulting in hopping like process. This might be what we witness in Na$_2$CsC%
$_{60}\,$ and this bears interesting similarities to the metal to
insulator transition observed at high temperature in manganites
with ferromagnetic metallic ground states. In these systems also
strong electronic correlations are coexisting with the possibility
of JTD of the doubly degenerate e$_g$ orbitals. It has been
proposed that the occurrence of JTD at high temperatures, which
forbids hopping in this case, adds to the double exchange
mechanism to trigger a transition to the high temperature
insulating phase \cite{Millis}.


\begin{references}
\bibitem{rq1}  The fact that the shift at the transition increases for $%
^{23} $Na and decreases for $^{133}$Cs definetely shows that it is the
hyperfine coupling that is changing and not the susceptibility.
\bibitem{rq2}  $^{23}$Na 1/T$_1$ would be a more sensitive test, but
quadrupolar relaxation probably dominates at high T and it cannot be used.
\end{references}
\end{document}